\documentclass[pra,twoside,showpacs,floatfix]{revtex4}
\usepackage{amsmath,amsfonts,graphicx}
\usepackage[amssymb]{SIunits}
\usepackage[active]{srcltx}
\usepackage{color}

\newcommand{\ket}[1]{|#1\rangle}
\newcommand{\bra}[1]{\langle#1|}
\newcommand{\bracket}[2]{\langle#1|#2\rangle}
 
\DeclareMathOperator{\id}{\openone}
\SetSymbolFont{symbols}{normal}{OMS}{cmsy}{m}{n}

\begin{document}

\bibliographystyle{apsrev}

\title{Relativistic Einstein-Podolsky-Rosen correlations for vector
  and tensor states} 

\author{Pawe{\l} Caban}\email{P.Caban@merlin.phys.uni.lodz.pl}
\author{Jakub Rembieli{\'n}ski}\email{jaremb@uni.lodz.pl}
\author{Marta W{\l}odarczyk}\email{marta.wlodarczyk@gmail.com}

\affiliation{Department of Theoretical Physics, University of Lodz\\
Pomorska 149/153, 90-236 {\L}{\'o}d{\'z}, Poland}

\date{\today}

\begin{abstract}
We calculate and investigate the relativistic correlation function for
bipartite systems of spin-$1/2$ in vector and spin-$1$ particles in
tensor states. We show that the relativistic correlation function,
which depends on particles momenta, may have local extrema. What is
more, the momentum dependance of the correlation functions for two
choices of relativistic spin operator may be significantly different. 
\end{abstract}

 \maketitle

\section{Introduction}
We have recently shown that for the scalar states the {relativistic}
Einstein-Podolsky-Rosen (EPR) correlation function, which depends on
the particle four-momenta, 
may have local extrema for certain fixed configurations  in
massive particle systems
\cite{ja2008, ja2009, Caban_2008_bosons_helicity,
  CDKO_2010_fermion_helicity, smolinski}.
Such extrema are present for bipartite systems of  both spin-$1/2$
and spin-$1$ particles, for two different 
choices of the relativistic spin operator. This {property} has not
been reported in any of the earlier papers {on the subject} 
\cite{ALMH2003, CR2005, CR2006, Czachor1997_1, CW2003, Czachor2005,
  czachor2010, FBHH2010, GA2002, GBA2003, LD2003, LY2004, KS2005,
  PT2004_1, PST2005, RS2002, SL2004, TU2003_1, Terno2003}. 

Moreover we have shown that in certain configurations   relativistic
quantum correlations  can be stronger than {in} the non-relativistic
{case}, leading to stronger violation of the Bell-type inequalities. 

The aim of the present paper is to extend our considerations on
non-scalar states. We present and discuss new results for the
correlation functions for vector states (for spin-$1/2$ particles) and
antisymmetric tensor states (for spin-$1$ particles). 

\section{Preliminaries}
In order to calculate correlations between spin degrees of freedom, we
need a properly defined relativistic spin operator. {The difficulty is
that} there does not  exist any unambiguous definition of such an
operator 
\cite{Czachor1997_1,CW2003,TU2003_1,Terno2003,LD2003,RS2002,LY2004,SL2004,CR2005,CR2006,czachor2010,bogolubov75}.
In calculations of relativistic EPR correlation functions one most
frequently uses the so called Newton-Wigner spin operator
\cite{newton-wigner} 
 \begin{equation}
 \label{eq:definicja_operatora_spinu}
 \hat{\boldsymbol{S}}_{NW}=\frac{1}{m}\left(\hat{\boldsymbol{W}}-
 \hat{W}^0\frac{\boldsymbol{\hat{P}}}{\hat{P}^0+m}\right),
 \end{equation}
or the operator related to the projection of the center-of-mass (c.m.)
spin operator on the direction $\boldsymbol{\omega}$ (see
e.g.~\cite{Czachor1997_1}) 
 \begin{equation}
 \label{spin_znormalizowany}
 \hat{S}_{cm}(\boldsymbol{\omega})= 
 \frac{\boldsymbol{\omega}\cdot\hat{\boldsymbol{W}}}{\sqrt{m^2+
 (\boldsymbol{\omega}\cdot\hat{\boldsymbol{P}})^2}}. 
 \end{equation}
For the discussion of this point see \cite{ja2009}. Below we perform
calculations for both choices of the spin operator. 

{The states, for which we calculate the relativistic correlation functions are defined as follows}.
The Hilbert space of a single particle of mass $m$ and spin $s$ is spanned by eigenvectors of the four-momentum operators $\hat{P}^{\mu}$:
\begin{equation}
    \label{eq:baza_pedowa}
    \hat{P}^{\mu}\ket{q,\sigma}=q^{\mu}\ket{q,\sigma},
\end{equation}
where $\sigma$ is the value of the spin projection on the $z$ axis;
$\sigma=\{-s,-s+1,\hdots ,s\}$, and $q^{\mu}=\left(q^0,\boldsymbol{q}\right)$.
We assume the covariant normalization of the states (\ref{eq:baza_pedowa}).

{In EPR-type experiments one deals with bipartite states. For the sake
of convenience we introduce the following notation for fermionic
(half-integer $s$) and bosonic (integer $s$) bipartite states: 
 \begin{gather}
  \label{eq:fermion}
  \ket{(k,\sigma),(p,\lambda)}_f=
  \frac{1}{\sqrt{2}}(\ket{k,\sigma}\otimes\ket{p,\lambda}-
  \ket{p,\lambda}\otimes\ket{k,\sigma}),\\
 \label{eq:boson}
  \ket{(k,\sigma),(p,\lambda)}_b= 
  \frac{1}{\sqrt{2}}(\ket{k,\sigma}\otimes\ket{p,\lambda}+
  \ket{p,\lambda}\otimes\ket{k,\sigma}),
 \end{gather}
respectively.

Classification of all bipartite states covariant with respect to the
Lorentz group action has been discussed in our earlier papers. For
vector bosons it was done in Ref.~\cite{ja2008} and for fermions in
Ref.~\cite{CR2006}. In particular for a bipartite system of spin-$1/2$
particles the vector state reads 
 \begin{equation}
 \label{eq:stan wektorowy}
 \ket{\varphi}=-i \varphi^{\mu}(\mathrm{v}^{\mathrm{T}}(k)
 \gamma^2\gamma^0\gamma_{\mu}\mathrm{v}(p))_{\sigma\lambda}
 \ket{(k,\sigma),(p,\lambda)}_f,
 \end{equation}
where $\gamma_{\mu}$ denotes the {Dirac} matrices (for the convention
used see \cite{CR2006}), $\ket{(k,\sigma),(p,\lambda)}_f$ is defined
by Eq.~(\ref{eq:fermion}) and the Dirac field amplitude
$\mathrm{v}(q)$ ($q=\{k,p\}$) is a $2\times 4$ matrix of the form
\cite{CR2006} 
  \begin{equation}
    \label{eq:v}
    \mathrm{v}(q)=\frac{1}{2\sqrt{1+\frac{q^0}{m}}}\left(
                                                     \begin{array}{c}
                                                       \left(\id+\frac{1}{m}q^{\mu}\sigma_{\mu}\right)\sigma_2 \\
                                                       \left(\id+\frac{1}{m}q^{\pi\mu}\sigma_{\mu}\right)\sigma_2 \\
                                                     \end{array}
                                                   \right),
  \end{equation}
  where $q^{\pi}=(q^0,-\boldsymbol{q})$, $\sigma_{\mu}=(\id,\sigma_i)$ and $\sigma_i$, $i=\{1,2,3\}$ are the Pauli matrices.

  {The tensor state for a bipartite system of spin-$1$ particles has the form 
    \label{seq:def_stanu_tensorowego}
    \begin{equation}
    \label{eq:stan_antysym_uv}
       \ket{\Phi}=\mathrm{T}_{\mu\nu}e^{\mu}_{\,\,\sigma}(k)e^{\nu}_{\,\,\lambda}(p)\ket{(k,\sigma),(p,\lambda)}_b,
    \end{equation}
  where $\ket{(k,\sigma),(p,\lambda)}_b$ is defined by Eq.~(\ref{eq:boson}) and the amplitude $e(q)$ ($q=\{k,p\}$) reads \cite{ja2008}}
    \begin{equation}
        \label{eq:e}
        e(q)\!=\!\left(\begin{array}{c}
         \tfrac{\boldsymbol{q}^T}{m}\\
         \hline
         \id+\tfrac{\boldsymbol{q}\otimes\boldsymbol{q}^T}{m(m+q^0)}
         \end{array}\right)V^{\mathrm{T}},
        \quad  V\!=\!\frac{1}{\sqrt{2}}\left(
        \begin{array}{ccc}
            -1 & i & 0 \\
            0 & 0 & \sqrt{2} \\
            1 & i & 0 \\
        \end{array}
        \right).\\
    \end{equation}
  As $T_{\mu\nu}$ we take the most general form of an antisymmetric tensor
    \begin{equation}
    \label{eq:tensor_asym}
        T_{\mu\nu}=\left(
        \begin{array}{c|c}
        0 & \alpha^j \\\hline
        -\alpha^i & \epsilon^{ijk}\beta^k \\
        \end{array}
        \right),
    \end{equation}
where $\boldsymbol{\alpha}$ and $\boldsymbol{\beta}$ are arbitrary complex vectors fixing the polarization of the state (\ref{eq:stan_antysym_uv}).

\section{Correlation functions}
Now let us {consider} two distant observers in the same inertial
frame of reference---Alice and Bob. Both share a pair of particles of
mass $m$ in one of the states (\ref{eq:stan wektorowy}) {or}
(\ref{eq:stan_antysym_uv}). {Alice measures the spin component of one
  particle along the direction $\boldsymbol{a}$ and Bob measures the
  spin component of the other particle along the direction
  $\boldsymbol{b}$. Their observables are $\boldsymbol{a}\cdot
  \hat{\boldsymbol{S}}_{NW}$ and $\boldsymbol{b}\cdot
  \hat{\boldsymbol{S}}_{NW}$ for (\ref{eq:definicja_operatora_spinu})
  and $\hat{S}_{cm}(\boldsymbol{a})$ and
  $\hat{S}_{cm}(\boldsymbol{b})$ for (\ref{spin_znormalizowany}),
  respectively.} The normalised correlation functions for an arbitrary
bipartite state $\ket{\psi}$ are {then given by} 
 \begin{equation}
 \label{eq:korelacje nw}
 \mathcal{C}_{NW}^{\psi(k,p)}(\boldsymbol{a},\boldsymbol{b})=
 \frac{\bra{\psi}(\boldsymbol{a}\cdot 
 \hat{\boldsymbol{S}}_{NW})(\boldsymbol{b}\cdot 
 \hat{\boldsymbol{S}}_{NW})\ket{\psi}}{s^2\bracket{\psi}{\psi}}
 \end{equation}
or
 \begin{equation}
 \label{eq:korelacje znormalizowany}
 \mathcal{C}_{cm}^{\psi(k,p)}(\boldsymbol{a},\boldsymbol{b})= 
 \frac{\bra{\psi}\hat{S}_{cm}(\boldsymbol{a})\hat{S}_{cm}(\boldsymbol{b})
 \ket{\psi}}{s^2\bracket{\psi}{\psi}},
 \end{equation}
respectively.

\subsection{The case of the system of spin-$1/2$ particles}

{In this subsection we calculate the correlation function for a
spin-$1/2$ system in a vector state 
for the c.m. spin operator [c.f.~Eq.~(\ref{spin_znormalizowany})]. The
corresponding correlation function for the 
Newton-Wigner spin operator
[Eq.~(\ref{eq:definicja_operatora_spinu})] has been calculated
elsewhere \cite{CR2006}.  In view of lengthy formulae in a general
case, we restrict our considerations to the case of the
center-of-mass frame (c.m. frame), that is to a frame, in which the
particles have opposite momenta ($\boldsymbol{p}=-\boldsymbol{k}$). 

The correlation function for the Newton-Wigner spin operator
(\ref{eq:definicja_operatora_spinu})  reads} \cite{CR2006} 
 \begin{multline}
 \label{eq:f_kor_cmf_vec_nw}
 C_{NW}^{{\varphi(k,k^{\pi})}}
 (x,\boldsymbol{a},\boldsymbol{b},\boldsymbol{\varphi})=
 \boldsymbol{a\cdot b}
 -\frac{(x+1)[(\boldsymbol{a}\cdot\boldsymbol{\varphi})
 (\boldsymbol{b}\cdot\boldsymbol{\varphi}^*)+
 (\boldsymbol{a}\cdot\boldsymbol{\varphi}^*)
 (\boldsymbol{b}\cdot\boldsymbol{\varphi})]}{(x+1)
 |\boldsymbol{\varphi}|^2
 -x|\boldsymbol{\varphi}\cdot\boldsymbol{n}|^2}\\
 -\frac{2x^2(\boldsymbol{a\cdot n})(\boldsymbol{b\cdot n})
 |\boldsymbol{\varphi}\cdot\boldsymbol{n}|^2}
 {(\sqrt{x+1}+1)^2
 \left[(x+1)|\boldsymbol{\varphi}|^2-
 x|\boldsymbol{\varphi}\cdot\boldsymbol{n}|^2\right]}
 +\frac{x\sqrt{x+1}}{\sqrt{x+1}+1}
 \frac{(\boldsymbol{a\cdot n})
 \left[(\boldsymbol{b}\cdot
 \boldsymbol{\varphi})(\boldsymbol{n}\cdot\boldsymbol{\varphi}^*)
 +(\boldsymbol{b}\cdot
 \boldsymbol{\varphi}^*)(\boldsymbol{n}\cdot\boldsymbol{\varphi})\right]}
 {(x+1)|\boldsymbol{\varphi}|^2
 -x|\boldsymbol{\varphi}\cdot\boldsymbol{n}|^2}\\
 +\frac{x\sqrt{x+1}}{\sqrt{x+1}+1}
 \frac{(\boldsymbol{b\cdot n})
 \left[(\boldsymbol{a}\cdot
 \boldsymbol{\varphi})(\boldsymbol{n}\cdot\boldsymbol{\varphi}^*)
 +(\boldsymbol{a}\cdot
 \boldsymbol{\varphi}^*)(\boldsymbol{n}\cdot\boldsymbol{\varphi})\right]}
 {(x+1)|\boldsymbol{\varphi}|^2
 -x|\boldsymbol{\varphi}\cdot\boldsymbol{n}|^2},
 \end{multline}
where $\boldsymbol{n}=\tfrac{\boldsymbol{k}}{|\boldsymbol{k}|}$, and
$x=\left(\tfrac{|\boldsymbol{k}|}{m}\right)^2$. 
For c.m. operator (\ref{spin_znormalizowany}) using
Eqs.~(\ref{eq:korelacje znormalizowany}) and (\ref{eq:stan wektorowy}), we get 
 \begin{multline}
 \label{eq:f_kor_cmf_vec_czachor}
 C_{cm}^{{{\varphi(k,k^{\pi})}}}
 (x,\boldsymbol{a},\boldsymbol{b},\boldsymbol{\varphi})=
 \frac{\boldsymbol{a\cdot b}+x(\boldsymbol{a\cdot n})
 (\boldsymbol{b\cdot n})}{\sqrt{1+x(\boldsymbol{a\cdot n})^2}
 \sqrt{1+x(\boldsymbol{b\cdot n})^2}} 
 -\frac{(x+1)[(\boldsymbol{a}\cdot\boldsymbol{\varphi})
  (\boldsymbol{b}\cdot\boldsymbol{\varphi}^*)
 +(\boldsymbol{a}\cdot\boldsymbol{\varphi}^*)
 (\boldsymbol{b}\cdot\boldsymbol{\varphi})]}{\sqrt{1+
 x(\boldsymbol{a\cdot n})^2}\sqrt{1+
 x(\boldsymbol{b\cdot n})^2}
 \left[(x+1)|\boldsymbol{\varphi}|^2-x|
 \boldsymbol{\varphi}\cdot\boldsymbol{n}|^2\right]}.
 \end{multline}

\begin{figure}
\centering
  \includegraphics[width=0.5\columnwidth]{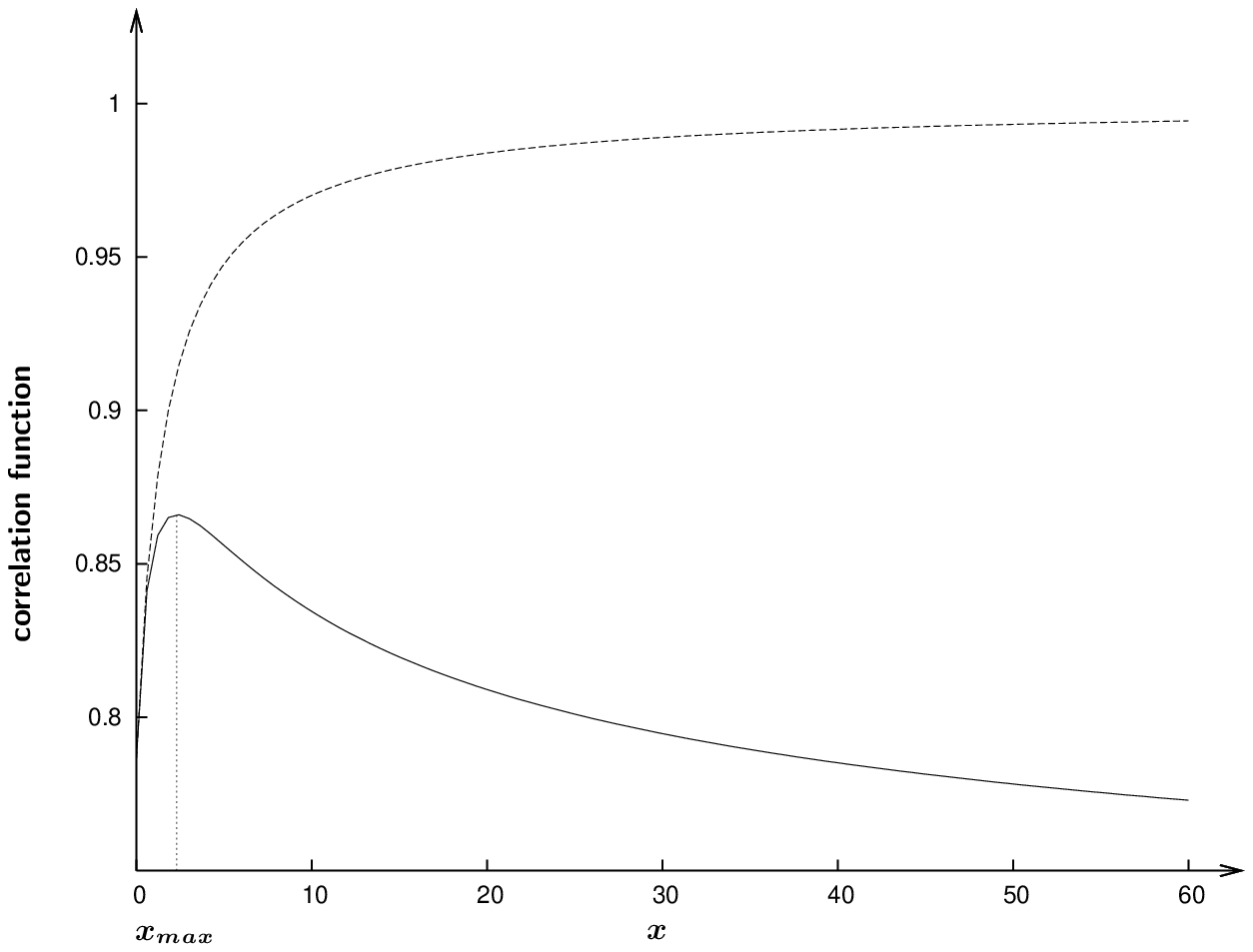}
  \caption{The plot shows dependance of
    $C_{NW}^{{{\varphi(k,k^{\pi})}}}
    (x,\boldsymbol{a},\boldsymbol{b},\boldsymbol{\varphi})$ 
    (solid line) and
    $C_{cm}^{{{\varphi(k,k^{\pi})}}}
    (x,\boldsymbol{a},\boldsymbol{b},\boldsymbol{\varphi})$ 
    (dashed line) on $x$ for $\boldsymbol{a\cdot n}=1$,
    $\boldsymbol{b\cdot n}=\boldsymbol{a\cdot b}={1}/{\sqrt{2}}$,
    $\boldsymbol{a}\cdot\boldsymbol{\varphi}=
    \boldsymbol{n}\cdot\boldsymbol{\varphi}={1}/{2}$ 
    and $\boldsymbol{b}\cdot\boldsymbol{\varphi}=-0.08$. The first one
    has maximum equal to $0.89$ at $x=2.30$ and the second one
    increases monotonically to $1$.}\label{1}
\end{figure}
Both functions may have local extrema for {specific values of
  arguments $\boldsymbol{a}$, $\boldsymbol{b}$, $\boldsymbol{n}$ and
  $\boldsymbol{\varphi}$. Moreover, the dependence of the correlation
  functions on $x$ differs for the spin operators
  (\ref{eq:definicja_operatora_spinu}) and
  (\ref{spin_znormalizowany}).  This is illustrated in
  Figs.~\ref{1}--\ref{3}, where the functions
  $C_{NW}^{{{\varphi(k,k^{\pi})}}}
  (x,\boldsymbol{a},\boldsymbol{b},\boldsymbol{\varphi})$ and
  $C_{cm}^{{{\varphi(k,k^{\pi})}}}
  (x,\boldsymbol{a},\boldsymbol{b},\boldsymbol{\varphi})$ of the
  argument $x$ are drawn for three sets of parameters
  $\boldsymbol{a}$, $\boldsymbol{b}$, $\boldsymbol{n}$ and
  $\boldsymbol{\varphi}$, as indicated in captions.  In Fig.~\ref{1},
  the function $C_{NW}^{{{\varphi(k,k^{\pi})}}}
  (x,\boldsymbol{a},\boldsymbol{b},\boldsymbol{\varphi})$ has a local
  maximum, while $C_{cm}^{{{\varphi(k,k^{\pi})}}}
  (x,\boldsymbol{a},\boldsymbol{b},\boldsymbol{\varphi})$
  monotonically approaches the value of $1$ for $x\rightarrow \infty$.
\begin{figure}
\centering
  \includegraphics[width=0.5\columnwidth]{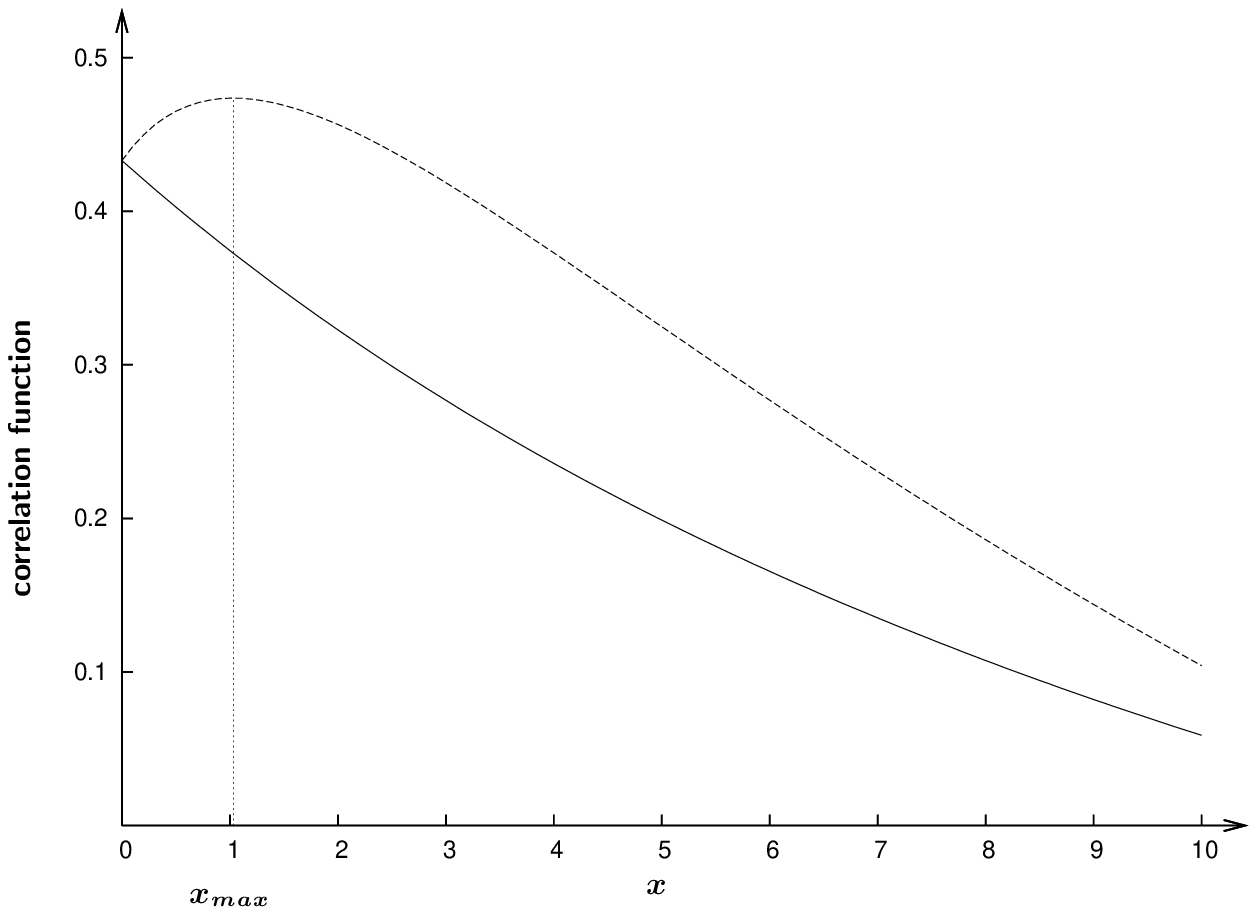}
  \caption{The plot shows dependance of
    $C_{NW}^{{{\varphi(k,k^{\pi})}}}
    (x,\boldsymbol{a},\boldsymbol{b},\boldsymbol{\varphi})$ 
    (solid line) and
    $C_{cm}^{{{\varphi(k,k^{\pi})}}}
    (x,\boldsymbol{a},\boldsymbol{b},\boldsymbol{\varphi})$ 
    (dashed line) on $x$ for $\boldsymbol{a\cdot n}=1$,
    $\boldsymbol{b\cdot n}=\boldsymbol{a\cdot b}=-{1}/{2}$,
    $\boldsymbol{a}\cdot\boldsymbol{\varphi}=
    \boldsymbol{n}\cdot\boldsymbol{\varphi}=-0.97$ 
    and $\boldsymbol{b}\cdot\boldsymbol{\varphi}=0.48$. The first one
    decreases monotonically to $-0.5$ in the ultra-relativistic limit,
    the second one has maximum equal to $0.47$ at $x=1.03$ and its
    ultra-relativistic limit is $-1$ (full anticorrelation).}\label{2}
\end{figure}
In Fig.~\ref{2} the function $C_{cm}^{{{\varphi(k,k^{\pi})}}}
(x,\boldsymbol{a},\boldsymbol{b},\boldsymbol{\varphi})$ has a maximum,
while $C_{NW}^{{{\varphi(k,k^{\pi})}}}
(x,\boldsymbol{a},\boldsymbol{b},\boldsymbol{\varphi})$ decreases
monotonically with increasing $x$. In Fig.~\ref{3} the function
$C_{NW}^{{{\varphi(k,k^{\pi})}}}
(x,\boldsymbol{a},\boldsymbol{b},\boldsymbol{\varphi})$ is constant
and equal $-0.5$, while $C_{cm}^{{{\varphi(k,k^{\pi})}}}
(x,\boldsymbol{a},\boldsymbol{b},\boldsymbol{\varphi})$ increases
monotonically with increasing $x$.}
\begin{figure}
\centering
  \includegraphics[width=0.5\columnwidth]{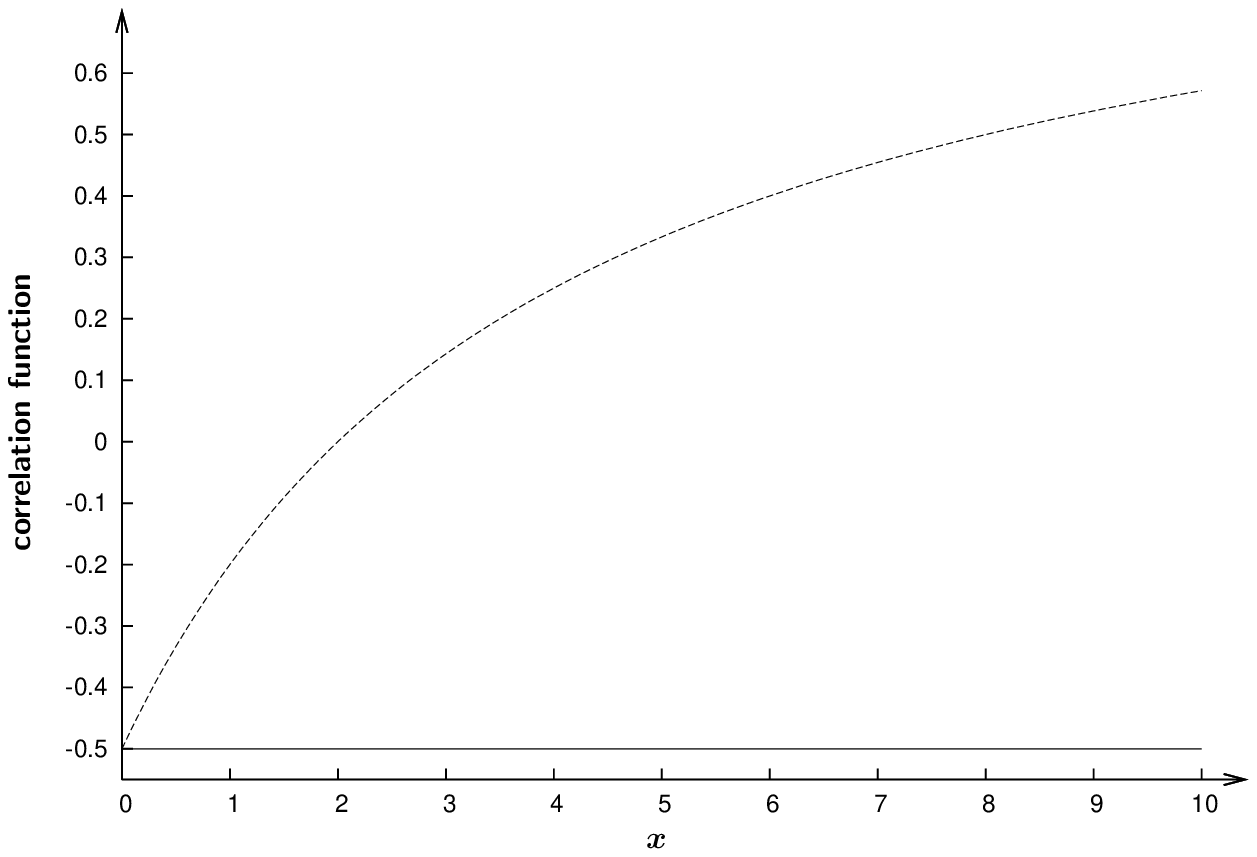}
  \caption{The plot shows dependance of
    $C_{NW}^{{{\varphi(k,k^{\pi})}}}
    (x,\boldsymbol{a},\boldsymbol{b},\boldsymbol{\varphi})$ 
    (solid line) and
    $C_{cm}^{{{\varphi(k,k^{\pi})}}}
    (x,\boldsymbol{a},\boldsymbol{b},\boldsymbol{\varphi})$ 
    (dashed line) on $x$ for $\boldsymbol{a\cdot n}=\boldsymbol{b\cdot
      n}={1}/{2}$, $\boldsymbol{a\cdot b}={1}/{4}$,
    $\boldsymbol{a}\cdot\boldsymbol{\varphi}=
    \boldsymbol{b}\cdot\boldsymbol{\varphi}={\sqrt{3}}/{2\sqrt{2}}$ 
    and $\boldsymbol{n}\cdot\boldsymbol{\varphi}=0$. The first
    function is constant and equal $-0.5$, the second one increases
    monotonically to reach $1$ (full correlation) in the
    ultra-relativistic limit.}\label{3}
\end{figure}

\subsection{The case of the system of the spin-$1$ particles}

{The correlation function of a bipartite vector boson system in the
  state (\ref{eq:stan_antysym_uv}) can be calculated by means of
  Eqs.~(\ref{eq:stan_antysym_uv}) and (\ref{eq:korelacje nw}). For
  simplicity we give the result in the c.m. frame.} 
\begin{widetext}
\begin{multline}
\label{eq:f_korelacji_asym_cmf}
    C_{NW}^{{{\Phi(k,k^{\pi})}}}(x,\boldsymbol{a},\boldsymbol{b},\boldsymbol{\alpha},\boldsymbol{\beta})=
    \frac{1}{2[x (\boldsymbol{\beta\cdot n})(\boldsymbol{\beta}^*\cdot\boldsymbol{n})-x(2x+1)(\boldsymbol{\alpha\cdot n})(\boldsymbol{\alpha}^*\cdot\boldsymbol{n})-(x+1)(\boldsymbol{\beta\cdot \beta}^*)-x (\boldsymbol{\alpha\cdot \alpha}^*)]}\times\\
    \times \left\{\sqrt{x}[(\boldsymbol{\alpha\cdot \beta}^*)(\boldsymbol{n}\cdot(\boldsymbol{a}\times\boldsymbol{b}))+
    (\boldsymbol{\beta}^*\cdot\boldsymbol{n})(\boldsymbol{\alpha}\cdot (\boldsymbol{a}\times\boldsymbol{b}))]\right.
       +(x+1)(\boldsymbol{a\cdot \beta})(\boldsymbol{b\cdot \beta}^*)\\
    -x(\boldsymbol{a}\cdot (\boldsymbol{\alpha}^*\times\boldsymbol{n}))(\boldsymbol{b}\cdot (\boldsymbol{\alpha}\times\boldsymbol{n}))
    +2(\sqrt{x+1}-1)^2(\boldsymbol{a\cdot n})(\boldsymbol{b\cdot n})(\boldsymbol{\beta\cdot n})(\boldsymbol{\beta}^*\cdot\boldsymbol{n})
    \\+\sqrt{x}(\sqrt{x+1}-1)[2(\boldsymbol{\alpha\cdot n})(\boldsymbol{\beta}^*\cdot\boldsymbol{n})(\boldsymbol{n}\cdot(\boldsymbol{a}\times\boldsymbol{b}))
    +(\boldsymbol{a\cdot n})(\boldsymbol{\beta}^*\cdot\boldsymbol{n})(\boldsymbol{b}\cdot (\boldsymbol{\alpha}\times\boldsymbol{n}))]
    \\-\left.
    \sqrt{x}(\sqrt{x+1}-1)[(\boldsymbol{a\cdot \beta}^*)(\boldsymbol{b}\cdot (\boldsymbol{\alpha}\times\boldsymbol{n}))+(\boldsymbol{b\cdot n})(\boldsymbol{\beta}^*\cdot\boldsymbol{n})(\boldsymbol{a}\cdot (\boldsymbol{\alpha}\times\boldsymbol{n}))
    -(\boldsymbol{b\cdot \beta}^*)(\boldsymbol{a}\cdot (\boldsymbol{\alpha}\times\boldsymbol{n}))]\right.\\
    -\sqrt{x+1}(\sqrt{x+1}-1)((\boldsymbol{a\cdot \beta})(\boldsymbol{b\cdot n})(\boldsymbol{\beta^*\cdot n})\left.+(\boldsymbol{b\cdot \beta})(\boldsymbol{a\cdot n})(\boldsymbol{\beta^*\cdot n}))
    \right\}+c.c.,\\
\end{multline}
\end{widetext}
where again $\boldsymbol{n}=\tfrac{\boldsymbol{k}}{|\boldsymbol{k}|}$,
$x=\left(\tfrac{|\boldsymbol{k}|}{m}\right)^2$, {and $c.c.$ stands for
  the complex conjugation}. For particular values of arguments, the
function may have local extrema.
{In particular, choosing the polarization where $\boldsymbol{\alpha}=\boldsymbol{0}$, function (\ref{eq:f_korelacji_asym_cmf}) reduces to the simpler form}
  \begin{multline}
  \label{eq:f_korelacji_asym_cmf_u0}
  C_{NW}^{{{\Phi(k,k^{\pi})}}}
  (x,\boldsymbol{a},\boldsymbol{b},\boldsymbol{\beta})=
  \frac{1}{x|\boldsymbol{\beta\cdot n}|^2-x-1}
  \left\{(x\!+\!1)(\boldsymbol{a\cdot \beta})(\boldsymbol{b\cdot \beta})
  \!+\!(\sqrt{x\!+\!1}\!-\!1)^2
 (\boldsymbol{a\cdot n})(\boldsymbol{b\cdot n})
 (\boldsymbol{\beta\cdot n})^2\right.\\
        -\left.({x+1}-\sqrt{x+1})(\boldsymbol{\beta\cdot
            n})[(\boldsymbol{a\cdot n})(\boldsymbol{b\cdot
            \beta})+(\boldsymbol{a\cdot \beta})(\boldsymbol{b\cdot
            n})]\right\}. 
\end{multline}
Now {taking into account that for any particle carrying four-momentum $\boldsymbol{q}$}
 \begin{equation}
 \label{eq:zwiazek}
 \hat{S}(\boldsymbol{\omega})=
 \frac{1}{\sqrt{m^2+(\boldsymbol{\omega}
 \cdot\boldsymbol{q})^2}}\left[m(\boldsymbol{\omega}\cdot
 \boldsymbol{\hat{S}}_{NW})\!+\!
 \hat{W}^0\frac{\boldsymbol{\omega}\cdot\boldsymbol{q}}{m+q^0}\right],
 \end{equation}
$\boldsymbol{\omega}=\{\boldsymbol{a},\boldsymbol{b}\}$, for the spin
operator (\ref{spin_znormalizowany}), we have 
 \begin{equation}
 \label{eq:f_korelacji_asym_cmf_czachor_u0}
 C_{cm}^{{{\Phi(k,k^{\pi})}}}
 (x,\boldsymbol{a},\boldsymbol{b},\boldsymbol{\beta})=
 \frac{(x+1)(\boldsymbol{a\cdot \beta})(\boldsymbol{b\cdot \beta})}
 {\sqrt{1+x(\boldsymbol{a\cdot n})^2}
 \sqrt{1+x(\boldsymbol{b\cdot n})^2}(x|\boldsymbol{\beta\cdot n}|^2-x-1)}.
 \end{equation}
\begin{figure}
\centering
  \includegraphics[width=0.5\columnwidth]{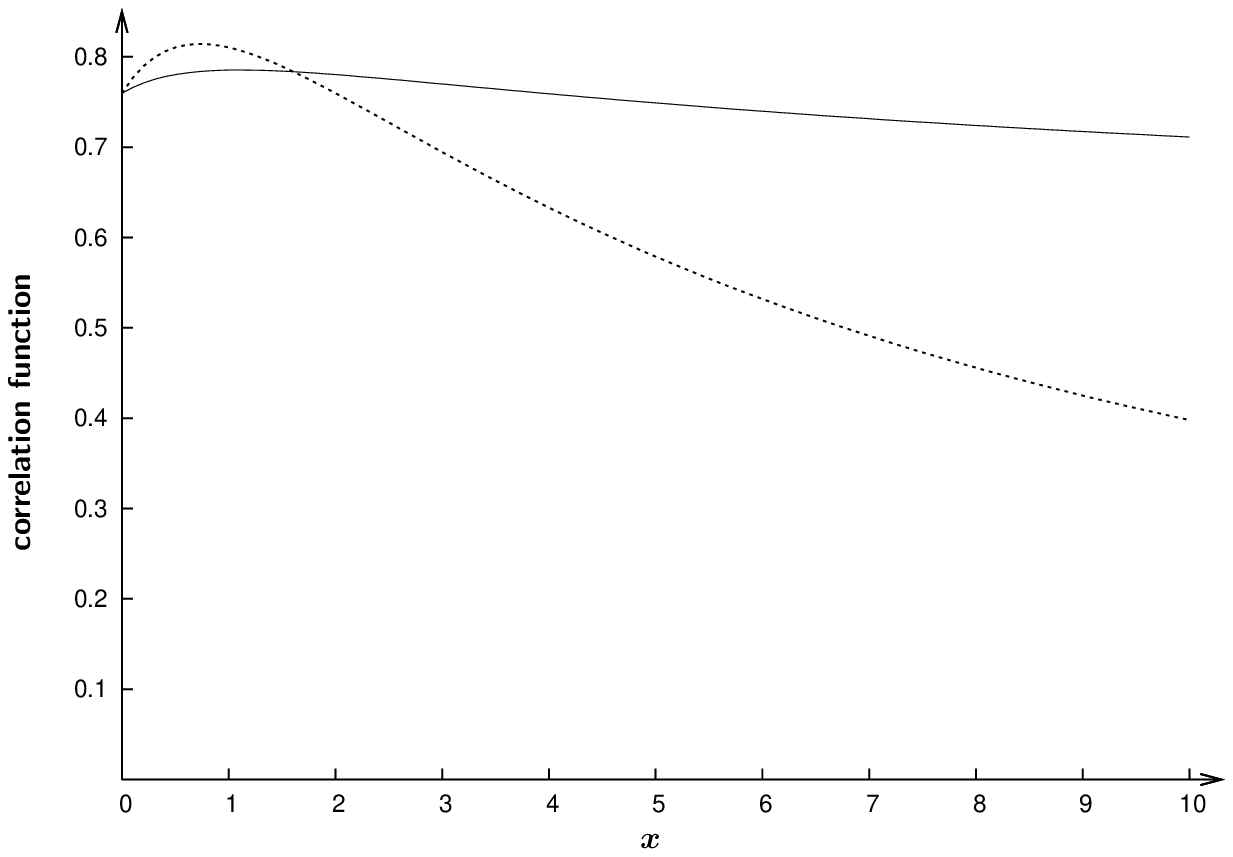}
  \caption{The plot shows dependence of
      $C_{NW}^{{{\Phi(k,k^{\pi})}}}
      (x,\boldsymbol{a},\boldsymbol{b},\boldsymbol{\beta})$ 
      (solid line) and
      $C_{cm}^{{{\Phi(k,k^{\pi})}}}
      (x,\boldsymbol{a},\boldsymbol{b},\boldsymbol{\beta})$ 
      (dashed line) for configuration $\boldsymbol{a\cdot
        n}=-\boldsymbol{b\cdot n}=-{1}/{2}$, $\boldsymbol{a\cdot
        b}=-0.90$, $\boldsymbol{a\cdot\beta}=-0.79$,
      $\boldsymbol{b\cdot\beta}=0.97$ and $\boldsymbol{\beta\cdot
        n}={1}/{\sqrt{2}}$. Both functions have local maxima --- the
      first one equal to $0,79$ for $x=0.81$, the second one $1.10$
      for $x=0.73$. Their ultra-relativistic limits are equal
      ${3}/{4\sqrt{2}}$ and $0$ respectively.}\label{19}
\end{figure}
{As in the case of spin-$1/2$ particles, the dependence of the
  correlation functions corresponding to the spin operators
  (\ref{eq:definicja_operatora_spinu}) and (\ref{spin_znormalizowany})
  may have significantly different dependences on the variable
  $x$. This is demonstrated in Figs.~\ref{19}--\ref{15}. 
\begin{figure}
\centering
  \includegraphics[width=0.5\columnwidth]{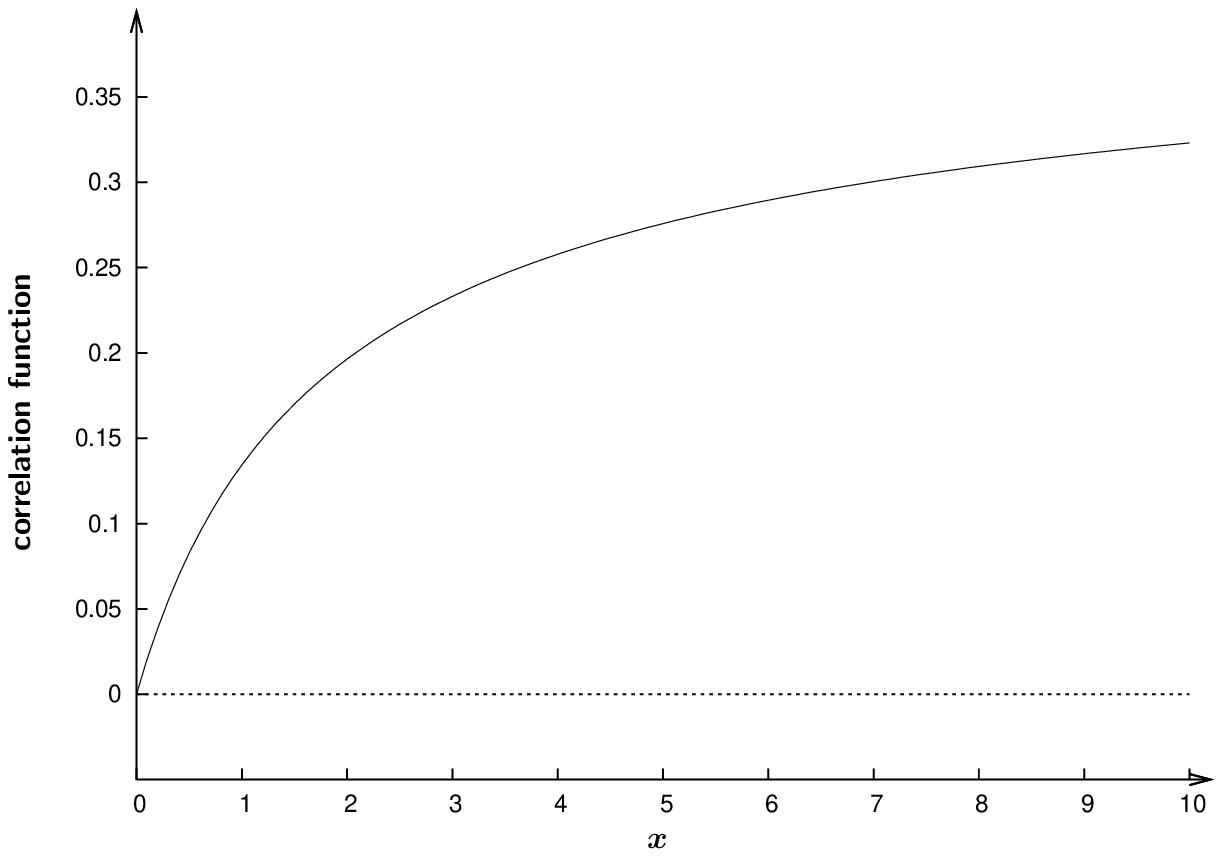}
  \caption{ The plot shows dependence of
    $C_{NW}^{{{\Phi(k,k^{\pi})}}}
    (x,\boldsymbol{a},\boldsymbol{b},\boldsymbol{\beta})$ 
    (solid line) and
    $C_{cm}^{{{\Phi(k,k^{\pi})}}}
    (x,\boldsymbol{a},\boldsymbol{b},\boldsymbol{\beta})$ 
    (dashed line) for configuration $\boldsymbol{a\cdot
      n}=\boldsymbol{\beta \cdot n}={1}/{2}$, $\boldsymbol{b\cdot
      n}={\sqrt{3}}/{2}$, $\boldsymbol{a\cdot
      b}=\boldsymbol{b\cdot\beta}=0$ i $\boldsymbol{a\cdot\beta}=1$.
    The function for (\ref{eq:definicja_operatora_spinu}) increases
    monotonically from $0$ to ${\sqrt{3}}/{4}$ in the
    ultra-relativistic limit.}\label{13}
\end{figure}
In Fig.~\ref{19} both functions have local maxima.
\begin{figure}
\centering
  \includegraphics[width=0.5\columnwidth]{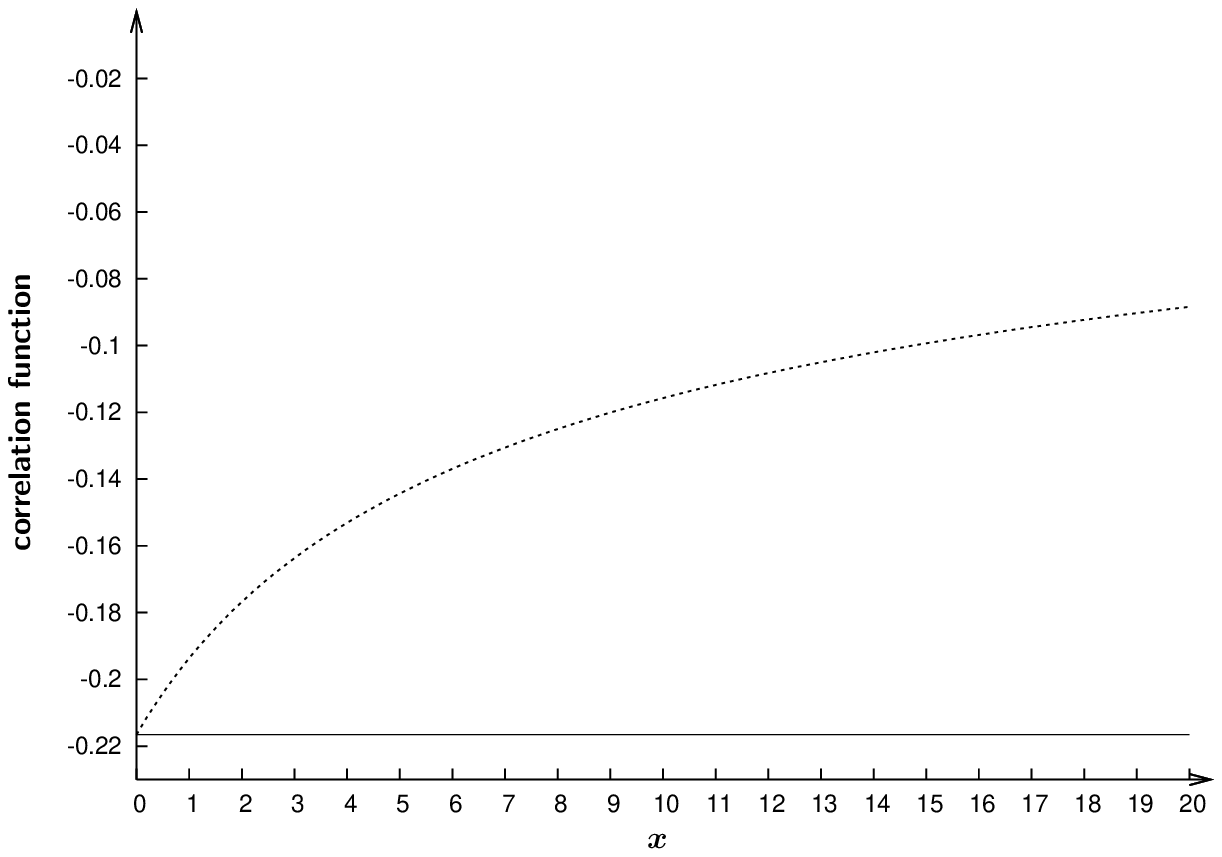}
  \caption{ The plot shows dependence of
    $C_{NW}^{{{\Phi(k,k^{\pi})}}}
    (x,\boldsymbol{a},\boldsymbol{b},\boldsymbol{\beta})$ 
    (solid line) and
    $C_{cm}^{{{\Phi(k,k^{\pi})}}}
    (x,\boldsymbol{a},\boldsymbol{b},\boldsymbol{\beta})$ 
    (dashed line) for configuration $\boldsymbol{a\cdot
      n}=-\boldsymbol{b\cdot\beta}={1}/{2}$, $\boldsymbol{b\cdot
      n}=\boldsymbol{\beta\cdot n}=0$, $\boldsymbol{a\cdot
      b}={\sqrt{3}}/{2}$ i $\boldsymbol{a\cdot\beta}=-{\sqrt{3}}/{4}$.
    The function for (\ref{eq:definicja_operatora_spinu}) is constant
    and equal to $-{\sqrt{3}}/{8}$, function for 
   (\ref{spin_znormalizowany}) monotonically increases from $-{\sqrt{3}}/{8}$
    to $0$ in the ultra-relativistic limit.}\label{14}
\end{figure}
In Figs.~\ref{13} and \ref{14} we have cases, for which one of the
functions is constant, while the other one is a monotonic function of
$x$. In Fig.~\ref{15} the function
$C_{NW}^{{{\Phi(k,k^{\pi})}}}
(x,\boldsymbol{a},\boldsymbol{b},\boldsymbol{\beta})$ 
is monotonically increasing to reach the value $0.75$ in the
ultra-relativistic limit, while the function
$C_{cm}^{{{\Phi(k,k^{\pi})}}}
(x,\boldsymbol{a},\boldsymbol{b},\boldsymbol{\beta})$ 
monotonically decreases to reach $0$ in infinity.} 
\begin{figure}
\centering
  \includegraphics[width=0.5\columnwidth]{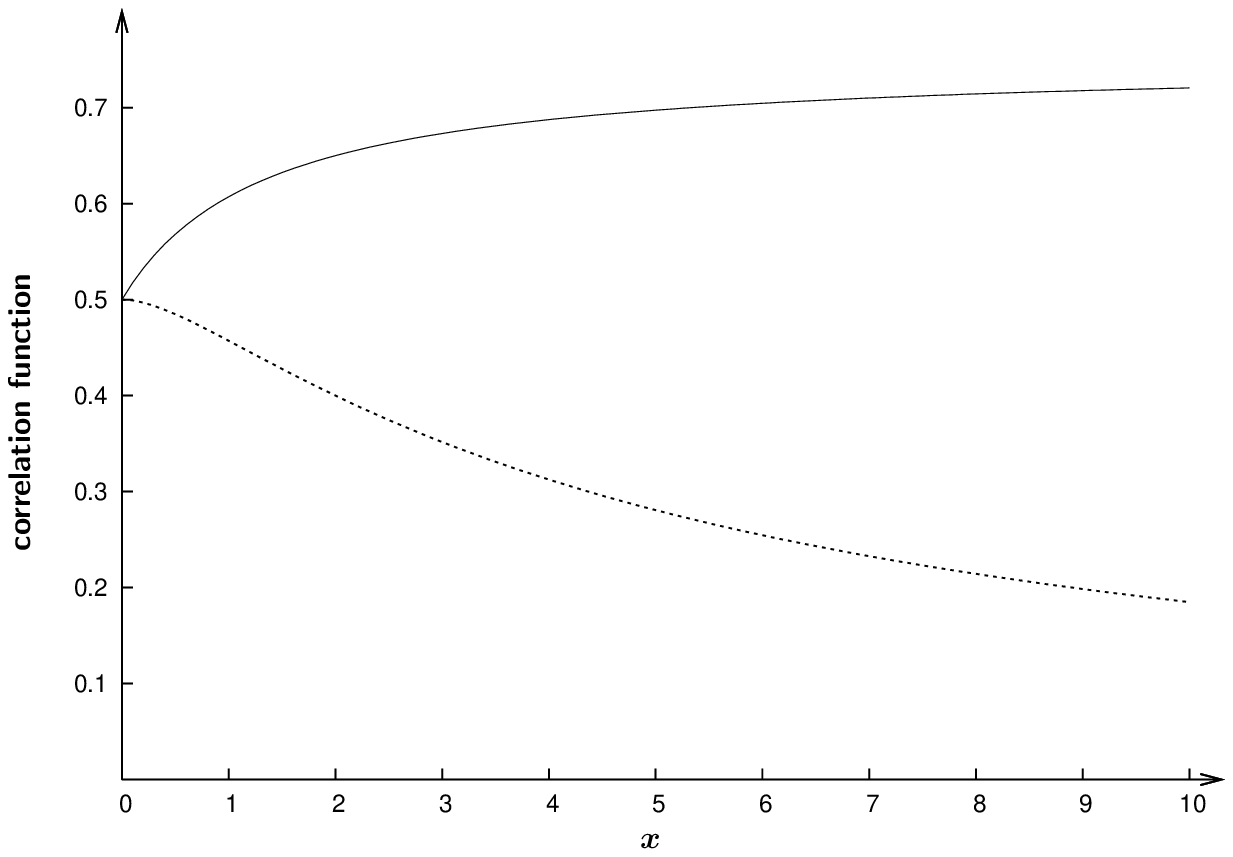}
  \caption{ The plot shows dependence of
    $C_{NW}^{{{\Phi(k,k^{\pi})}}}
    (x,\boldsymbol{a},\boldsymbol{b},\boldsymbol{\beta})$ 
    (solid line) and
    $C_{cm}^{{{\Phi(k,k^{\pi})}}}
    (x,\boldsymbol{a},\boldsymbol{b},\boldsymbol{\beta})$ 
    (dashed line) for configuration $\boldsymbol{a\cdot
      n}=\boldsymbol{b\cdot n}=\boldsymbol{\beta\cdot n}={1}/{2}$,
    $\boldsymbol{a\cdot b}=\boldsymbol{b\cdot\beta}=-{1}/{2}$ i
    $\boldsymbol{a\cdot\beta}=1$. For (\ref{spin_znormalizowany}), the
    correlations fade away with the increase of particles momenta and
    for (\ref{eq:definicja_operatora_spinu}) the particles become more
    correlated. in the ultra-relativistic limit correlation function
    for (\ref{eq:definicja_operatora_spinu}) operator equals
    $0.75$.}\label{15}
\end{figure}

\section{Conclusions}
In our previous papers \cite{ja2008,ja2009}, we have shown that the
relativistic correlation functions in bipartite systems of
spin-$1/2$ and spin-$1$ particles in a singlet state may have local
extrema in terms of particles momentum for particular values of
arguments. 
In this paper we have shown that a similar property appears for the
correlation functions in states which are not Lorentz singlets. 

The correlation functions have been calculated for two different
choices of the relativistic spin operator, and in both cases the
functions showed qualitatively similar behaviour. Nevertheless for
particular values of parameters, the behaviour of the correlation
functions for different spin operators could be quantitatively
different. This feature could be used for testing which  relativistic
spin operator gives predictions which are in better agreement with the
experimental data. 

\begin{acknowledgments}
The authors are grateful to Jacek Ciborowski for interesting
discussion.
This work has been supported by the University of Lodz. JR and PC have
been supported by the 
Ministry of Science and Higher Education under the contract N N202
103738. MW has been supported by the
Ministry of Science and Higher Education under the contract N N202
288938.
\end{acknowledgments}

%\bibliography{quant-ph1,odzbyszka,rel_entanglement}

\end{document}